# Модель Поттса на дереве Кэли: новый класс гиббсовских мер

М.М.Рахматуллаев, Ж.Д.Дехконов

Для модели Поттса на деревьях Кэли дается очень широкий класс новых мер Гиббса. Дается обзор известных гиббсовских мер модели Поттса на деревьях и эти меры сравниваются с нашими новыми мерами.

**Ключевые слова:** дерево Кэли, модель Поттса, мера Гиббса, слабо периодические меры Гиббса.

## 1 Введение

Решение задач, возникающих при исследовании термодинамических свойств физических и биологических систем, в основном проводится в рамках теории мер Гиббса. Мера Гиббса – это фундаментальное понятие, определяющее вероятность микроскопического состояния данной физической системы (определенной конкретным Гамильтонианом). Известно, что каждой мере Гиббса сопоставляется одна фаза физической системы, и если мера Гиббса не единственна, то говорят, что существует фазовый переход. Для достаточно широкого класса гамильтонианов известно, что множество всех предельных мер Гиббса (соответствующих данному гамильтониану) образует непустое выпуклое компактное подмножество в множестве всех вероятностных мер (см., например, [1]-[3]) и каждая точка этого выпуклого множества однозначно разлагается по его крайним точкам.

Хорошо известная модель ближайших соседей Поттса на дереве Кэли все еще предлагает новые интересные явления (см., например, [5]-[12], [15]-[21] для недавних результатов). В работе [13] для модели Изинга найдены множество новых мер Гиббса. В этой работе аналогично работе [13] мы расширяем набор известных мер Гиббса для модели Поттса.

Работа имеет следующую структуру. В разделе 2 даются необходимые определения и постановка задачи. Раздел 3 посвящен изучению новых мер Гиббса. В разделе 4 дается сравнение наших мер с известными мерами Гиббса.

## 2 Определения и постановка задачи

Дерево Кэли $T^k$ порядка $k \geq 1-$ бесконечное дерево, т.е. граф без циклов, из каждой вершины которого выходит ровно $k+1$ ребер. Пусть $T^k = (V, L, i)$, где $V -$ есть множество вершин $T^k$, $L -$ его множество ребер, и $i -$ функция инцидентности, сопоставляющая каждому ребру $l \in L$ его концевые точки

$x, y \in V$. Если $i(l) = \{x, y\}$, то $x$ и $y$ называются *ближайшими соседями вершин* и обозначается $l = \langle x, y \rangle$.

Расстояние $d(x, y), x, y \in V$ на дереве Кэли определяется формулой
$d(x, y) = \min\{d \mid \exists x = x_0, x_1, \ldots, x_{d-1}, x_d = y \in V$ такие, что $\langle x_0, x_1 \rangle, \ldots, \langle x_{d-1}, x_d \rangle\}$.

Для фиксированного $x^0 \in V$ обозначим $W_n = \{x \in V \mid d(x, x^0) = n\}$,
$V_n = \{x \in V \mid d(x, x^0) \leq n\}$, $L_n = \{l = \langle x, y \rangle \in L \mid x, y \in V_n\}$.

Для $x \in W_n$ положим $S(x) = \{y \in W_{n+1} : d(x, y) = 1\}$.

Известно, что существует взаимно однозначное соответствие между множеством $V$ вершин дерева Кэли порядка $k \geq 1$ и группой $G_k$, являющейся свободным произведением $k+1$ циклических групп второго порядка с образующими $a_1, a_2, \ldots, a_{k+1}$, соответственно (см. [4]).

Мы рассмотрим модель, где спиновые переменные принимают значения из множества $\Phi = \{1, 2, \ldots, q\}$, $q \geq 2$ и расположены на вершинах дерева. Тогда *конфигурация* $\sigma$ на $V$ определяется как функция $x \in V \to \sigma(x) \in \Phi$; множество всех конфигураций совпадает с $\Omega = \Phi^V$.

Гамильтониан модели Поттса определяется как

$$H(\sigma) = -J \sum_{\langle x, y \rangle \in L} \delta_{\sigma(x)\sigma(y)}, \tag{1}$$

где $J \in \mathbb{R}$, $x, y -$ ближайшие соседи и $\delta_{ij} -$ символ Кронекера:

$$\delta_{ij} = \begin{cases} 0, \text{ если } i \neq j, \\ 1, \text{ если } i = j. \end{cases}$$

Определим конечномерное распределение вероятностной меры $\mu_n$ в объеме $V_n$ как

$$\mu_n(\sigma_n) = Z_n^{-1} \exp\left\{-\beta H_n(\sigma_n) + \sum_{x \in W_n} h_{\sigma(x), x}\right\}, \tag{2}$$

где $\beta = 1/T$, $T > 0 -$ температура, $Z_n^{-1} -$ нормирующий множитель, $\{h_x = (h_{1,x}, \ldots, h_{q,x}) \in \mathbb{R}^q, x \in V\}$ совокупность векторов и

$$H_n(\sigma_n) = -J \sum_{\langle x, y \rangle \in L_n} \delta_{\sigma(x)\sigma(y)}.$$

Говорят, что вероятностное распределение (2) согласованное, если для всех $n \geq 1$ и $\sigma_{n-1} \in \Phi^{V_{n-1}}$

$$\sum_{\omega_n \in \Phi^{W_n}} \mu_n(\sigma_{n-1} \vee \omega_n) = \mu_{n-1}(\sigma_{n-1}), \tag{3}$$

Здесь $\sigma_{n-1} \vee \omega_n$ – есть объединение конфигураций. В этом случае, существует единственная мера $\mu$ на $\Phi^V$ такая, что для всех $n$ и $\sigma_n \in \Phi^{V_n}$

$$\mu(\{\sigma|_{V_n} = \sigma_n\}) = \mu_n(\sigma_n).$$

Такая мера называется расщепленной гиббсовской мерой, соответствующей гамильтониану (1) и векторнозначной функции $h_x, x \in V$.

Следующее утверждение описывает условие на $h_x$, обеспечивающее согласованность $\mu_n(\sigma_n)$.

**Теорема 1** *(см.[4]) Вероятностное распределение $\mu_n(\sigma_n), n=1,2,...$ в (2) является согласованным тогда и только тогда, когда для любого $x \in V$ имеет место следующее*

$$\widetilde{h}_x = \sum_{y \in S(x)} F(\widetilde{h}_y, \theta), \tag{4}$$

где функция $F: h = (h_1,...,h_{q-1}) \in \mathbb{R}^{q-1} \to F(h,\theta) = (F_1,...,F_{q-1}) \in \mathbb{R}^{q-1}$ определяется как:

$$F_i = \ln\left(\frac{(\theta-1)\exp^{h_i} + \sum_{j=1}^{q-1}\exp^{h_j} + 1}{\theta + \sum_{j=1}^{q-1}\exp^{h_j}}\right),$$

$\theta = \exp(J\beta)$, $S(x)$ – множество прямых потомков точки $x$ и $\widetilde{h}_x = (\widetilde{h}_{1,x},...,\widetilde{h}_{q-1,x})$ с условием

$$\widetilde{h}_{i,x} = h_{i,x} - h_{q,x}, i = 1,...,q-1.$$

Каждому решению $\widetilde{h}_x$ функционального уравнения (4) соответствует одна мера Гиббса и наоборот.

А именно, для любого граничного условия $\widetilde{h}_x$, удовлетворяющего функциональному уравнению (4), существует единственная мера Гиббса, которая называется расщепляющей мерой Гиббса (РМГ). Более того, согласно [1, Теорема 12.6] любая крайняя мера Гиббса является РМГ; поскольку каждая мера Гиббса может быть представлена как крайними. Проблема полного описания мер Гиббса модели Поттса сводится к описанию класса РМГ. Поэтому в данной статье мы рассматриваем только РМГ и опускаем слово расщепление.

Целью работы является описание множества новых гиббсовских мер для модели Поттса на дереве Кэли.

## 3 Новые меры Гиббса.

В этой работе мы рассматриваем полубесконечное дерево. А именно, корень $x^0$ имеет $k$ ближайших соседей.

В этом пункте мы находим новые решения функционального уравнения (4) при $q=3$. Рассмотрим следующую матрицу

$$M = \begin{pmatrix} a & b \\ c & d \end{pmatrix},$$

где $a,b,c,d$ целые неотрицательные числа и
$$a+b=k, c+d=k. \qquad (5)$$

Эта матрица определяет того, что сколько раз присутствуют векторные значения $\bar{h}=(h_1,h_2)$ и $\bar{l}=(l_1,l_2)$ в множестве $S(x)$ для каждого $h_x \in \{\bar{h},\bar{l}\}$. Граничные условия (т.е совокупность векторов) $h=\{h_x, x\in V\}$, определяются следующим образом:

(i) если в вершине $x$ имеем $h_x = \bar{h}$, то функцию $h_y$ принимающую векторное значение на каждой вершине $y \in S(x)$ определим по следующему правилу:

$$\begin{cases} \bar{h}, \text{ на } a \text{ вершинах } S(x); \\ \bar{l}, \text{ на } b \text{ вершинах } S(x). \end{cases}$$

(ii) если в вершине $x$ имеем $h_x = \bar{l}$, то функция имеет значения:

$$\begin{cases} \bar{h}, \text{ на } c \text{ вершинах } S(x); \\ \bar{l}, \text{ на } d \text{ вершинах } S(x). \end{cases}$$

Легко видеть, что граничные условия $h_x$ (т.е. совокупность векторов) приведенное выше конструкции удовлетворяют функциональному уравнению (4), если векторы $\bar{h}=(h_1,h_2)$ и $\bar{l}=(l_1,l_2)$ удовлетворяют следующей системе уравнений:

$$\begin{cases} h_1 = a\ln\dfrac{\theta\exp(h_1)+\exp(h_2)+1}{\theta+\exp(h_1)+\exp(h_2)}+b\ln\dfrac{\theta\exp(l_1)+\exp(l_2)+1}{\theta+\exp(l_1)+\exp(l_2)}, \\ h_2 = a\ln\dfrac{\theta\exp(h_2)+\exp(h_1)+1}{\theta+\exp(h_1)+\exp(h_2)}+b\ln\dfrac{\theta\exp(l_2)+\exp(l_1)+1}{\theta+\exp(l_1)+\exp(l_2)}, \\ l_1 = c\ln\dfrac{\theta\exp(h_1)+\exp(h_2)+1}{\theta+\exp(h_1)+\exp(h_2)}+d\ln\dfrac{\theta\exp(l_1)+\exp(l_2)+1}{\theta+\exp(l_1)+\exp(l_2)}, \\ l_2 = c\ln\dfrac{\theta\exp(h_2)+\exp(h_1)+1}{\theta+\exp(h_1)+\exp(h_2)}+d\ln\dfrac{\theta\exp(l_2)+\exp(l_1)+1}{\theta+\exp(l_1)+\exp(l_2)}, \end{cases} \quad (6)$$

где $a,b,c,d$ задаются в матрице $M$.

Рассмотрим отображение $W:\mathbb{R}^4 \to \mathbb{R}^4$, определенное в следующем виде:

$$\begin{cases} h'_1 = a\ln\dfrac{\theta\exp(h_1)+\exp(h_2)+1}{\theta+\exp(h_1)+\exp(h_2)}+b\ln\dfrac{\theta\exp(l_1)+\exp(l_2)+1}{\theta+\exp(l_1)+\exp(l_2)}, \\ h'_2 = a\ln\dfrac{\theta\exp(h_2)+\exp(h_1)+1}{\theta+\exp(h_1)+\exp(h_2)}+b\ln\dfrac{\theta\exp(l_2)+\exp(l_1)+1}{\theta+\exp(l_1)+\exp(l_2)}, \\ l'_1 = c\ln\dfrac{\theta\exp(h_1)+\exp(h_2)+1}{\theta+\exp(h_1)+\exp(h_2)}+d\ln\dfrac{\theta\exp(l_1)+\exp(l_2)+1}{\theta+\exp(l_1)+\exp(l_2)}, \\ l'_2 = c\ln\dfrac{\theta\exp(h_2)+\exp(h_1)+1}{\theta+\exp(h_1)+\exp(h_2)}+d\ln(\dfrac{\theta\exp(l_2)+\exp(l_1)+1}{\theta+\exp(l_1)+\exp(l_2)}. \end{cases} \quad (7)$$

Система (6) эквивалентна системе уравнений $h=W(h)$. Легко доказывается следующая

**Лемма 1** *Отображение $W$ имеет инвариантные множества следующих видов:*

$$I_1 = \{h=(h_1,h_2,l_1,l_2)\in R^4: h_2=l_2=0\},$$

$$I_2 = \{h=(h_1,h_2,l_1,l_2)\in R^4: h_1=l_1=0\},$$

$$I_3 = \{h=(h_1,h_2,l_1,l_2)\in R^4: h_1=h_2, l_1=l_2\}.$$

i) Система уравнений (6) на $I_1$ имеет следующий вид:

$$\begin{cases} h_1 = af_\theta(h_1)+bf_\theta(l_1), \\ l_1 = df_\theta(l_1)+cf_\theta(h_1), \end{cases} \quad (8)$$

где $f_\theta(x)=\ln\dfrac{\theta\exp(x)+2}{\theta+\exp(x)+1}$.

Легко доказывается следующая

**Лемма 2** *Функция $f_\theta(x)$ обладает следующими свойствами:*
1. Если $\theta > 1$, то $f_\theta(x)$ возрастающая, если $\theta < 1$, то $f_\theta(x)$ убывающая.
2. Если $\theta > 1$, то $a < f_\theta(x) < A$. Если $\theta < 1$, то $A < f_\theta(x) < a$, где $a = \ln\frac{2}{\theta+1}, A = \ln\theta.$
3. $f_\theta(0) = 0$.
4. $\frac{d}{dx}f_\theta(0) = \frac{\theta-1}{\theta+2}$.

Теперь изучим систему уравнений (8)

**Лемма 3** *Независимо от параметров система уравнений (8) имеет решение (0,0), и если $|(bc-ad)(\frac{\theta-1}{\theta+2})^2 + (a+d)(\frac{\theta-1}{\theta+2})| > 1$, то существует не менее трех различных решений $(0,0)$, $(h_1^{(1)}, l_1^{(1)}), (h_1^{(2)}, l_1^{(2)})$, где $h_1^{(1)}, l_1^{(1)} > 0, h_1^{(2)}, l_1^{(2)} < 0.$*

*Proof.* Систему (8) рассмотрим в следующих частных случаях:
(1) Пусть $a = c$. Из условия леммы следует:

$$|(a+d)(\frac{\theta-1}{\theta+2})| > 1. \qquad (9)$$

Известно, что $a + d = k$, а так как $\theta > 1$ неравенство (9) имеет следующий вид:

$$\theta > \frac{k+2}{k-1}. \qquad (10)$$

При $a = c$ из системы (8) получим следующую систему:

$$\begin{cases} h_1 = af_\theta(h_1) + bf_\theta(l_1) \\ l_1 = bf_\theta(l_1) + af_\theta(h_1) \end{cases} \qquad (11)$$

из (11) следует что $h_1 = l_1$. В результате получим:

$$h_1 = kf_\theta(h_1), \qquad (12)$$

из [9] нам известно что уравнение (12) при условии $\theta > \theta_{cr} = \dfrac{k+2}{k-1}$ имеет не менее трех решений.

Значит, при выполнении условия леммы система (8) имеет не менее трех решений.

(2) Пусть $a = 0$. В этом случае, из первого уравнения системы (8) имеем равенство $h_1 = bf_\theta(l_1)$, где $b = k$. Тогда из второго уравнения получаем следующее равенство:

$$l_1 = g_\theta(l_1) = cf_\theta(kf_\theta(l_1)) + df_\theta(l_1). \tag{13}$$

Используя лемму 2, можно заключить, что $g_\theta(0) = 0$, $\dfrac{d}{dx}g_\theta(0) = ck\left(\dfrac{\theta-1}{\theta+2}\right)^2 + d\left(\dfrac{\theta-1}{\theta+2}\right)$ и $g_\theta(l_1)$ является ограниченной функцией от $l_1$. Кроме того, если $\left|\dfrac{d}{dx}g_\theta(0)\right| > 1$ (т.е 0 - нестабильная фиксированная точка $g$) то существует достаточно малая окрестность $(-\varepsilon, \varepsilon)$ у нуля: такая, что $g_\theta(l_1) < l_1$, при $l_1 \in (-\varepsilon, 0)$ и $g_\theta(l_1) > l_1$, при $l_1 \in (0, \varepsilon)$. Для $l_1 \in (0, \varepsilon)$ итерации $g_\theta^n(l_1)$ остаются больше нуля, монотонно возрастает и следовательно, сходятся к пределу $l_1^{(1)} > 0$ который является решением (13). Однако $l_1^{(1)} > 0$, так как точка 0 неустойчиво. Для $l_1 \in (-\varepsilon, 0)$ итерации $g_\theta^n(l_1)$ остаются меньше нуля, монотонно убывает и следовательно, сходятся к пределу $l_1^{(2)} < 0$ который является решением (13). Однако $l_1^{(2)} < 0$, так как точка 0 неустойчиво. Таким образом,

- Если выполняются условия $\left|\dfrac{d}{dx}g_\theta(0)\right| > 1$, то система (8) имеет не менее трех решений которые имеют следующий вид:

$(0,0), (kf_\theta(l_1^{(1)}), l_1^{(1)}), (kf_\theta(l_1^{(2)}), l_1^{(2)})$

(3) Пусть $b = 0$. Тогда из системы уравнений (8) имеем систему:

$$\begin{cases} h_1 = kf_\theta(h_1), \\ l_1 = df_\theta(l_1) + cf_\theta(h_1). \end{cases} \tag{14}$$

Первое уравнение этой системы имеет решение $h_1 = 0$ независимо от переменного. Используя это решение, второе уравнение системы (14) приводится к

$$l_1 = df_\theta(l_1), \tag{15}$$

которое имеет не менее трех решений в $\theta > \dfrac{d+2}{d-1}$ [9].

- Итак, при выполнении условия $b=0$, т.е. $\theta > \dfrac{d+2}{d-1}$, система (8) имеет не менее трех решений.

(4) Пусть $ab \neq 0$. В этом случае, из первого уравнения (8) получим равенство $f_\theta(l_1) = \dfrac{1}{b}(h_1 - af_\theta(h_1))$.

Из этого, используя второе уравнение, имеем следующее:
$$l_1 = \phi_\theta(h_1) = cf_\theta(h_1) + \dfrac{d}{b}(h_1 - af_\theta(h_1)) = \dfrac{1}{b}((bc-ad)f_\theta(h_1) + dh_1).$$

В результате, первое уравнение (8) можно выразить через формулу:
$$h_1 = \psi_\theta(h_1) = af_\theta(h_1) + bf_\theta(\dfrac{1}{b}((bc-ad)f_\theta(h_1) + dh_1)). \tag{16}$$

Ясно что, в $\psi_\theta(0) = 0$ и при выполнении аналогичных равенств во (2)-случае уравнение (16) может иметь не менее трех решений, при выполнении условия $|\dfrac{d}{dx}\psi_\theta(0)| = |(bc-ad)(\dfrac{\theta-1}{\theta+2})^2 + (a+d)(\dfrac{\theta-1}{\theta+2})| > 1$.

Итак (8) имеет по крайней мере три решения:
$(h_i, \phi_\theta(h_i)), i = 1, 2, 3$.

**Замечание 1** *При выполнении условия леммы 3, система (8) также может иметь более трех решений. Например, при $b = 0$ получается $h_1 = 0$ – решение первое уравнение системы уравнений (14), в общем случае это уравнение может иметь другое решение в условиях леммы.*

ii) Система уравнений (6) на $I_2$ имеет следующий вид:
$$\begin{cases} h_2 = az_\theta(h_2) + bz_\theta(l_2), \\ l_2 = cz_\theta(h_2) + dz_\theta(l_2), \end{cases} \tag{17}$$
где $z_\theta(x) = \ln \dfrac{\theta \exp(x) + 2}{\theta + \exp(x) + 1}$.

Теперь изучим систему уравнений (17)

**Лемма 4** *Независимо от параметров система уравнений (17) имеет решение (0,0), и если $|(bc-ad)(\dfrac{\theta-1}{\theta+2})^2 + (a+d)(\dfrac{\theta-1}{\theta+2})| > 1$ то существует не менее трех различных решений $(0,0), (h_2^{(1)}, l_2^{(1)}), (h_2^{(2)}, l_2^{(2)})$, где $h_2^{(1)}, l_2^{(1)} > 0, h_2^{(2)}, l_2^{(2)} < 0$.*

*Proof.* Эта лемма доказывается аналогично доказательству леммы 3

iii) Система уравнений (6) на $I_3$ имеет следующий вид:
$$\begin{cases} h_1 = a\varphi_\theta(h_1) + b\varphi_\theta(l_1), \\ l_1 = c\varphi_\theta(h_1) + d\varphi_\theta(l_1), \end{cases} \quad (18)$$

где $\varphi_\theta(x) = \ln\dfrac{(\theta+1)\exp(x)+1}{\theta+2\exp(x)}$.

Легко доказывается следующая

**Лемма 5** *Функция $\varphi_\theta(x)$ обладает следующими свойствами:*

1. Если $\theta > 1$, то $\varphi_\theta(x)$ возрастающая, если $\theta < 1$, то $\varphi_\theta(x)$ убывающая

2. Если $\theta > 1$, то $a < \varphi_\theta(x) < A$. Если $\theta < 1$, то $A < \varphi_\theta(x) < a$. где $a = \ln\dfrac{1}{\theta}, A = \ln\dfrac{\theta+1}{2}$.

3. $\varphi_\theta(0) = 0$.

4. $\dfrac{d}{dx}\varphi_\theta(0) = \dfrac{\theta-1}{\theta+2}, \ 0 < \dfrac{d}{dx}\varphi_\theta(x) < \dfrac{\theta-1}{\theta+2}, \theta > 1$.

Отметим, что системы уравнений (18) и (8) одинаковы, но функции $f_\theta(x)$ и $\varphi_\theta(x)$ разные. Аналогично рассуждая легко доказать следующую

**Лемма 6** *Независимо от параметров система уравнений (18) имеет решение (0,0) и если $|(bc-ad)(\dfrac{\theta-1}{\theta+2})^2 + (a+d)(\dfrac{\theta-1}{\theta+2})| > 1$, то существует не менее трех различных решений $(0,0), \ (h_1^{(1)}, l_1^{(1)}), (h_1^{(2)}, l_1^{(2)}),$ где $h_1^{(1)}, l_1^{(1)} > 0, h_1^{(2)}, l_1^{(2)} < 0.$*

**Замечание 2** *При выполнении условия леммы, система (18) также может иметь более трех решений. Например, при $b = 0$ получается решение $h_1 = 0$ решение первого уравнения системы уравнений (18), в общем случае это уравнение может иметь другое решение в условиях леммы.*

Объединяя леммы 3, 4 и 6 получим следующую теорему

**Теорема 2** *Если параметры $a, b, c, d$ и $\theta$ удовлетворяют следующим условиям:*

$$|(bc-ad)(\frac{\theta-1}{\theta+2})^2 + (a+d)(\frac{\theta-1}{\theta+2})|>1,$$

тогда для модели Поттса существуют не менее семи различных мер Гиббса, которые соответствуют решениям системы уравнений (6) вида (0,0,0,0), $(h_1^{(1)},0,l_1^{(1)},0)$, $(h_1^{(2)},0,l_1^{(2)},0)$, $(0,h_2^{(1)},0,l_2^{(1)})$, $(0,h_2^{(2)},0,l_2^{(2)})$, $(h_1^{(1)},h_1^{(1)},l_1^{(1)},l_1^{(1)})$, $(h_1^{(2)},h_1^{(2)},l_1^{(2)},l_1^{(2)})$.

Отметим, что мера Гиббса, соответствующая решению (0,0,0,0) системы уравнений (6) является трансляцион-инвариантным. Меры Гиббса, соответствующие совокупности векторов, удовлетворяющих функциональному уравнению (4) $\bar{h}=(h_1,h_2)$ и $\bar{l}=(l_1,l_2)$, полученной в теореме2 обозначим через $\mu_{h,l}$.

## 4 Сравнение меры $\mu_{h,l}$ с известными мерами Гиббса

*Трансляционно-инвариантные меры.* (см., например, [9, 10]). Такие меры соответствуют $h_x \equiv h = (h_1,...,h_{q-1}) \in \mathbb{R}^{q-1}$, т.е. постоянным функциям. Эти меры являются частными случаями наших мер, упомянутых в теоремы 2, которые могут быть получены при $a=c$ и $\bar{h}=\bar{l}$. В этом случае (6) имеет вид

$$h = kf(h). \qquad (19)$$

Из [9] известно, что уравнение (19) имеет три различных решения $h=0, h_1, h_2 (h_1>0, h_2<0)$, если $\theta > \theta_{cr} = \frac{k+2}{k-1}$. Отметим, что при выполнение условия теоремы 2 также выполняется условие $\theta > \theta_{cr} = \frac{k+2}{k-1}$.

*ART конструкция.* (см. [14],[15]) Пусть $\tilde{h}$ – граничное условие, удовлетворяющее (4) на $T^{k_0}$. Для $k \geq k_0$ определим следующее граничное условие на $T^k$:

$$\tilde{h}_x = \begin{cases} h, & x \in V^{k_0} \\ 0, & x \in V^k \setminus V^{k_0}, \end{cases} \qquad (20)$$

где $V^k$ обозначает множество вершин $T^k$. А именно, к каждой вершине $V^{k_0}$ добавляется $k-k_0$ последователей с нулевым значением граничного условия. Очевидно, что граничное условие $\tilde{h}$ удовлетворяет (4).

В случае, когда $h$ трансляционно-инвариантна на $T^{k_0}$, соответствующие меры этой конструкции могут быть получены по теореме 2 при $a=k_0, b=k-k_0$ и $l = 0$ (см. пример на Рис. 1).

Но в случае, когда $\tilde{h}$ не является трансляционно-инвариантной, меры ART на дерево Кэли $T^k$ не совпадают с мерами теоремы 2.

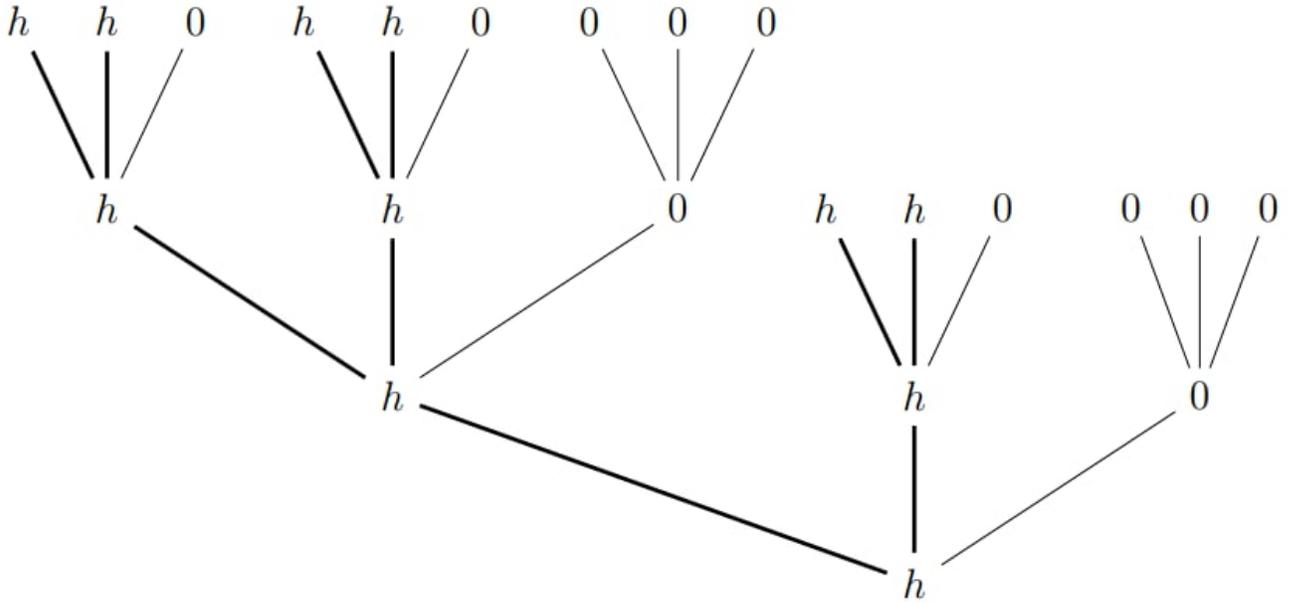

Рис 1: Это пример функции $\tilde{h}_x$ на вершинах дерева Кэли порядка 3. Это случай с условием $a=2, b=1, c=0, d=3, l=0$. На рисунке показаны два возможные правила размещения значений функции $\tilde{h}_x$ на множестве $S(x)$, т.е. на бесконечное $\tilde{h}_x$ количество элементов из $S(x)$ на $\tilde{h}_x$ ставят два $h$ и одна ноль, а на бесконечное количество $0$ ставят только нули.

($k_0$) –*трансляционно-инвариантная мера Гиббса.* (см. [15],[16]) Пусть $q=3, k_0 < k$ и векторы $\bar{h}_1, \bar{h}_2$ соответствуют трансляционно-инвариантному меру Гиббса на дереве Кэли порядка $k_0$. Теперь с помощью вектора $\bar{h}_1, \bar{h}_2$ на дереве Кэли порядка $k$ строим совокупность векторов $h_x : V \to \mathbb{R}$ следующим образом:

($l$) Пусть $k = p + q + k_0, p, q \in \mathbb{N}$. Если на вершине $x \in V$ имеем $h_x = \bar{h}_1$, то вершинам $S_{p+k_0}(x)$ сопоставляем вектора $h_x = \bar{h}_1$, остальным вершинам $S_q(x)$ сопоставляем вектора $h_x = \bar{h}_2$. Если на вершине $x \in V$ имеем $h_x = \bar{h}_2$, то вершинам $S_{q+k_0}(x)$ сопоставляем вектора $h_x = \bar{h}_2$, остальным вершинам $S_p(x)$ сопоставляем вектора $h_x = \bar{h}_1$ (см. Рис. 2).

**Определение 1** *Для модели Поттса мера, соответствующая*

совокупности векторов, построенных по правилам (l) называется
($k_0$)–трансляционно-инвариантной мерой Гиббса.

В случае $k_0 = 2$ и для фиксированных $\bar{h}_1, \bar{h}_2$ в работе [15] доказано существование не менее двух (2)–трансляционно-инвариантных мер Гиббса. Заметим, что эти (2)-трансляционно-инвариантные меры Гиббса являются частными случаями мер из теоремы 2 которые могут быт получены при $a = p + 2, b = q, d = p, c = q + 2$.

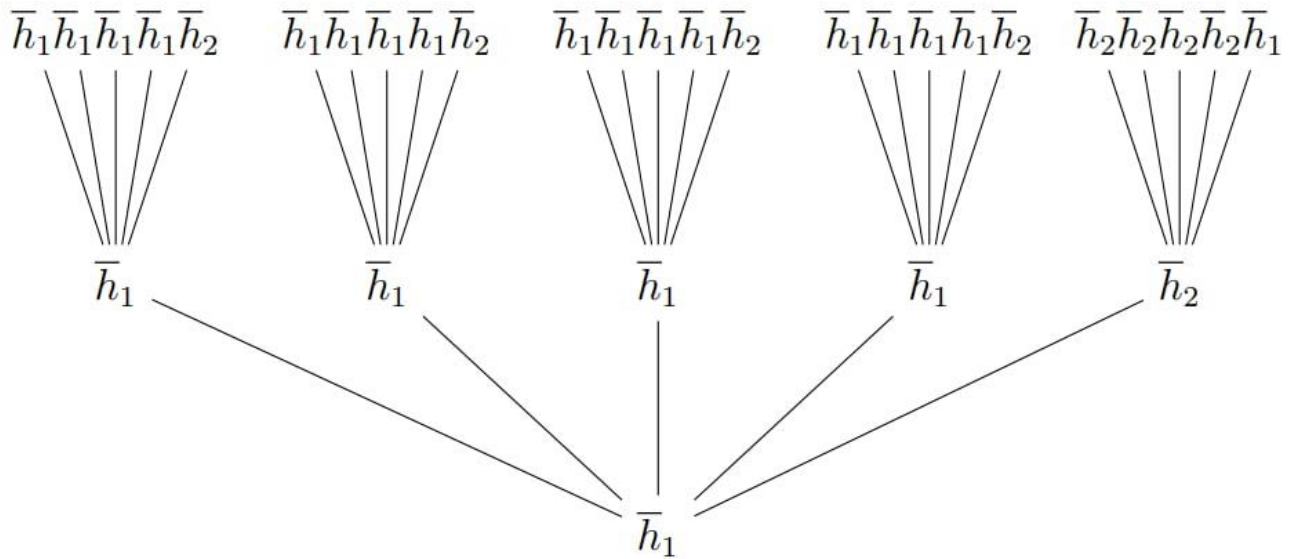

Рис 2: Приведем пример (3)–трансляционно-инвариантной функции $h_x$ на вершинах дерева Кэли порядка 5. Здесь $p = 1, q = 1$ и предполагается, что векторы $\bar{h}_1, \bar{h}_2$ соответствуют трансляционно-инвариантной мере Гиббса на дереве Кэли порядка три.

*Периодические меры Гиббса.* Пусть $G_k / \hat{G}_k = \{H_1,...,H_r\}-$ фактор-группа, где $\hat{G}_k$ – нормальная подгруппа индекса $r \geq 1$.

**Определение 2** *Совокупность векторов $h = \{h_x : x \in G_k\}$ называется $\hat{G}$-периодической, если $h_x = h_i$ при $x \in H_i$ для любых $x \in G_k$.*

Пусть $G_k^{(2)} = \{x \in G_k : длина\, слова\, x\, четна\}$, тогда $G_k^{(2)}$-периодическая совокупность векторов имеет следующий вид:

$$h_x = \begin{cases} h, & \text{если } x \in G_k^{(2)}, \\ l, & \text{если } x \in G_k \setminus G_k^{(2)}. \end{cases} \qquad (21)$$

Пусть $h = (h_1, h_2)$, $l = (l_1, l_2)$, тогда из (4) имеем

$$\begin{cases} h_i = k\ln\dfrac{(\theta-1)\exp(l_i) + \sum\limits_{i=1}^{2}\exp(l_j) + 1}{\sum\limits_{i=1}^{2}\exp(l_j) + \theta}, \\ l_i = k\ln\dfrac{(\theta-1)\exp(h_i) + \sum\limits_{i=1}^{2}\exp(h_j) + 1}{\sum\limits_{i=1}^{2}\exp(h_j) + \theta}. \end{cases} \quad i = 1, 2. \qquad (22)$$

Введем обозначения $\exp(h_i) = x_i$, $exp(l_i) = y_i$. Затем мы можем переписать последнюю систему уравнений для $i = 1, 2$ как

$$\begin{cases} x_i = \left(\dfrac{(\theta-1)y_i + \sum\limits_{i=1}^{2}(y_j) + 1}{\sum\limits_{i=1}^{2}(y_j) + \theta}\right)^k, \\ y_i = \left(\dfrac{(\theta-1)x_i + \sum\limits_{i=1}^{2}(x_j) + 1}{\sum\limits_{i=1}^{2}(x_j) + \theta}\right)^k. \end{cases} \qquad (23)$$

В случае $k = 2, q = 3$ и $J < 0$ доказано, что все $G_k^{(2)}$-периодические меры Гиббса на основе инварианта $I = \{(x_1, x_2, y_1, y_2) \in R^4 : x_1 = x_2, y_1 = y_2\}$ являются трансляционно-инвариантны (см.[7]).

В случае $k \geq 1, q = 3$ и $J > 0$ доказано, что все $G_k^{(2)}$-периодические меры Гиббса являются трансляционно-инвариантными (см.[7]).

В случае $k \geq 3, q = 3$ доказано, что система уравнений (23) на $I_0 = \{z = (u, v) \in R^2 : x_1 = x, y_1 = y; x_2 = y_2 = 1\}$ имеет не менее три решения при $0 < \theta < \dfrac{k-2}{k+1}$ (см.[8]).

Заметим, что эти меры являются частными случаями мер из теоремы 2, которые соответствуют решению системы уравнений (6) при $a = 0, b = k$ и $c = k, d = 0$. (См. Рис. 3, при $k = 3$).

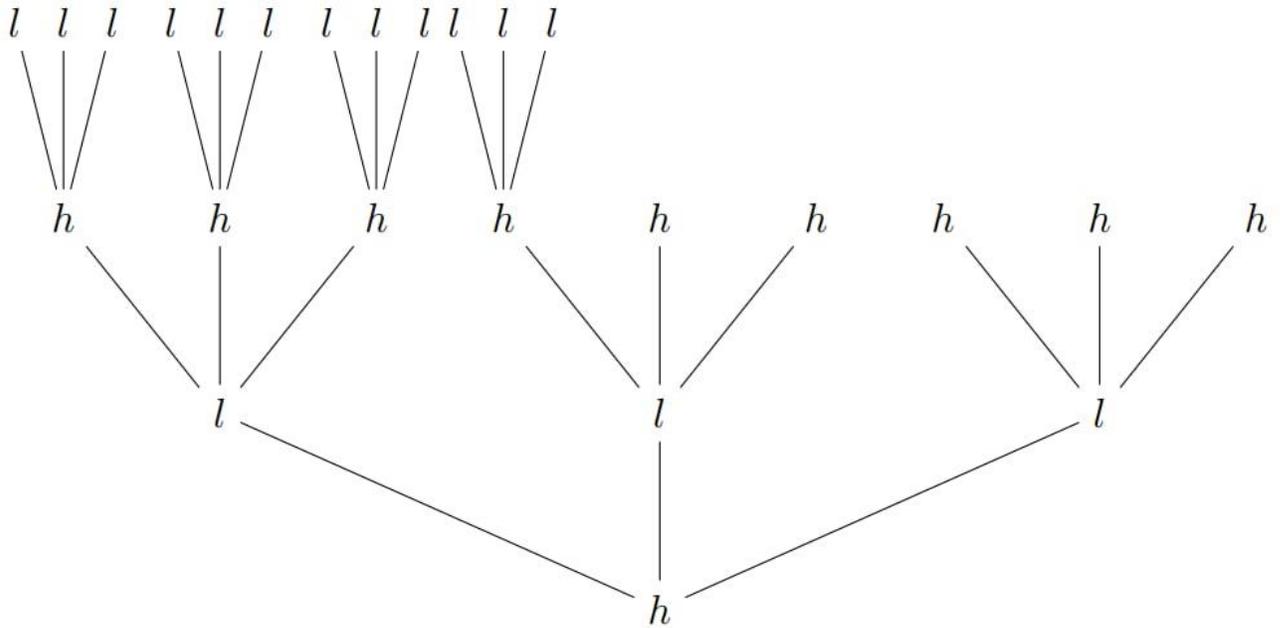

Рис 3: Приведем пример два-периодической функции $h_x$ на вершинах дерева Кэли порядка 3. Здесь $a=0, b=3, c=3, d=0$.

*Слабо периодические меры Гиббса.* Следуя [17-19], напомним понятие слабо периодической меры Гиббса.

**Определение 3** *Совокупность векторов* $h=\{h_x, x \in G_k\}$ *называется* $G_k$-*слабо периодической, если* $h_x = h_{ij}$ *при* $x \in H_i, x_\downarrow \in H_j$, *для любых* $x \in G_k$, *где через* $x_\downarrow$ *обозначим предок* $x$.

Слабо периодические граничные условия $h$ совпадают с периодическими, если $h_x$ не зависит от $x_\downarrow$.

Напомним результаты, известные для нормальной подгруппы индекса два. Заметим, что любая подгруппа индекса два имеет вид:

$$H_A = \{x \in G_k : \sum_{i \in A} w_x(a_i) - \text{четно}\}, \qquad (24)$$

где $\varnothing \neq A \subseteq N_k = \{1, 2, \ldots, k+1\}$, а $\omega_x(a_i)$ -количество $a_i$ в слове $x \in G_k$. Рассмотрим $A \neq N_k$.

Пусть $G_k/H_A = \{H_0, H_1\}$–фактор-группа, где $H_0 = H_A, H_1 = G_k \setminus H_A$. Тогда в силу (4) $H_A$-слабо периодические совокупности векторов

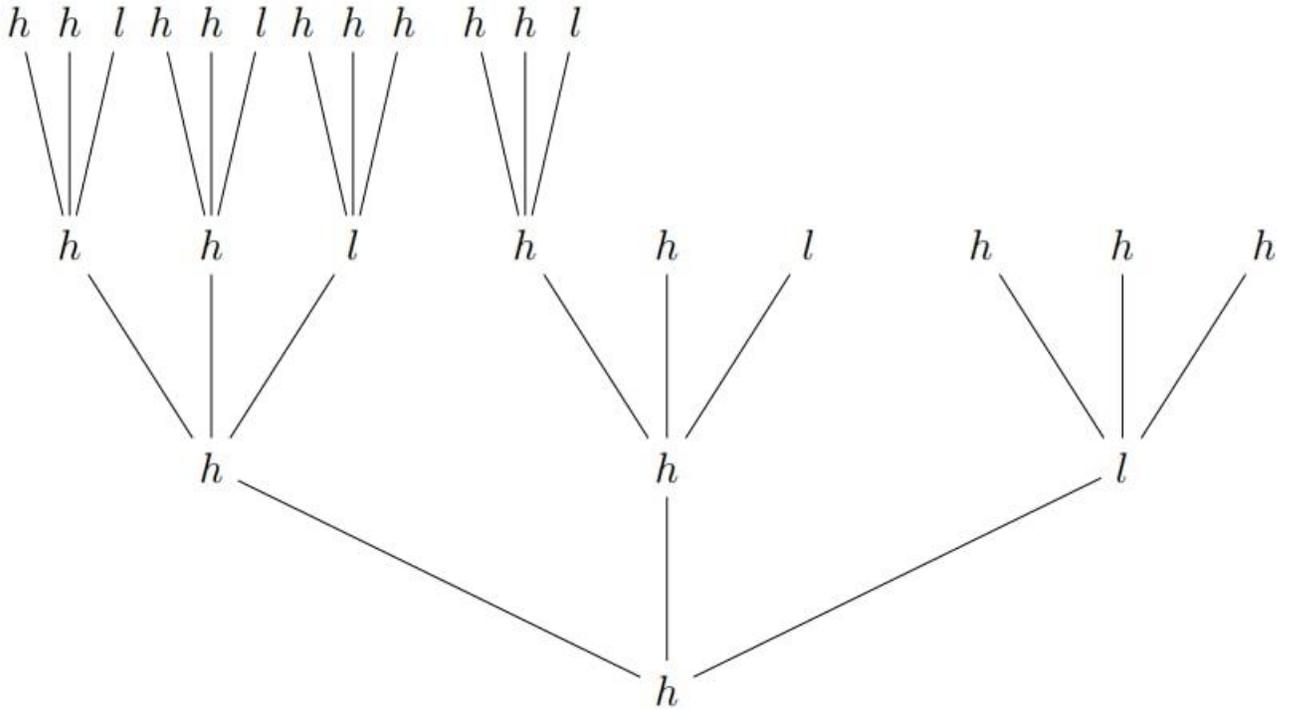

Рис 4: Приведем пример слабо-периодической функции $h_x$ на вершинах дерева Кэли порядка 3. Здесь $a=2, b=1, c=3, d=0$.

$$h_x = \begin{cases} h_1, & x \in H_0, x_\downarrow \in H_0, \\ h_2, & x \in H_0, x_\downarrow \in H_1, \\ h_3, & x \in H_1, x_\downarrow \in H_0, \\ h_4, & x \in H_1, x_\downarrow \in H_1, \end{cases} \qquad (25)$$

удовлетворяют следующей системе уравнений:

$$\begin{cases} h_1 = (k-|A|)f_\theta(h_1) + |A|f_\theta(h_2), \\ h_2 = (|A|-1)f_\theta(h_3) + (k+1-|A|)f_\theta(h_4), \\ h_3 = (|A|-1)f_\theta(h_2) + (k+1-|A|)f_\theta(h_1), \\ h_4 = (k-|A|)f_\theta(h_4) + |A|f_\theta(h_3). \end{cases} \qquad (26)$$

Очевидно, что относительно оператора $W : \mathbb{R}^4 \to \mathbb{R}^4$ инвариантами, определяемыми правой частью (26) являются следующие множества:

$I_1 = \{h \in \mathbb{R}^4 : h_1 = h_2 = h_3 = h_4\}$, $I_2 = \{h \in \mathbb{R}^4 : h_1 = h_4; h_2 = h_3\}$,

Заметим, что это
-меры, соответствующие решениям на $I_1$, являющиеся трансляционно-инвариантными, т.е. частными случаями мер, указанных в теореме 2.

-меры, соответствующие решениям на $I_2$, слабо периодические, которые совпадает с мерами, приведенными в теореме 2, при $a=k-|A|, b=|A|, c=k+1-|A|, d=|A|-1$. (См. Рис. 4, при $k=3, |A|=1$)

При этом система (26) решена только в случаях $|A|=1$ и $|A|=k$ (см. [17-19]). Таким образом, теорема 2 дает, в частности, новые слабо периодические меры Гиббса.

## Литература

# Сведения об авторах

1. Рахматуллаев Музаффар Мухаммаджанович, Доктор физ.-мат. наук, Заведующий отделом института математики им. В. И. Романовского АН РУз, Ташкент, Узбекистан.

адрес: Институт математики им. В. И. Романовского АН РУз, Узбекистан, 100174, Ташкент, ул. Университетская, 7.

e-mail: mrahmatullaev@rambler.ru

2. Дехконов Жасурбек Дилмурод угли, докторант Андижанского государственного университета.

адрес: Андижанский государственный университет, ул. УНИВЕРСИТЕТСКАЯ, 129, 170100, Андижан, Узбекистан.

e-mail: dehqonov.jasur@bk.ru